\documentclass[conference]{IEEEtran}
\IEEEoverridecommandlockouts
\usepackage{tikz}
\usepackage[utf8]{inputenc}
\usepackage{subcaption}  % For subfigure environment

\usepackage{pgfplots}
\usepackage{amsmath}
\usepackage{subcaption}
\usepackage{caption}
\usepackage{cite}
\usepackage{amsmath,amssymb,amsfonts}
\usepackage{algorithmic}
\usepackage{graphicx}
\usepackage{textcomp}
\usepackage{xcolor}
\def\BibTeX{{\rm B\kern-.05em{\sc i\kern-.025em b}\kern-.08em
    T\kern-.1667em\lower.7ex\hbox{E}\kern-.125emX}}

\begin{document}

\title{Enhanced Scalability of Horseshoe-and-Spur
Networks by Exploiting Hollow-Core fiber
%\thanks{The authors acknowledge support from the EU Horizon Europe project ECSTATIC.}
}

\author{\IEEEauthorblockN{Mohammad M. Hosseini}
%\orcidlink{0000-0003-1843-3536}
\IEEEauthorblockA{\textit{Nokia, Optical Networks}\\
Munich, Germany \\
mohammad.hosseini@nokia.com}

\and
\IEEEauthorblockN{João Pedro}
\IEEEauthorblockA{\textit{Nokia, Optical Networks}, 
Carnaxide, Portugal\\
\textit{Instituto de Telecomunicações, Instituto Superior Técnico}\\
Lisboa, Portugal\\
joao.pedro@nokia.com}

\and
\IEEEauthorblockN{Antonio Napoli}
%\orcidlink{0000-0002-9264-9274}
\IEEEauthorblockA{\textit{Nokia, Optical Networks}\\
Munich, Germany \\
antonio.napoli@nokia.com}
}
\maketitle
\begin{abstract}
The convergence of metro and access networks into unified optical infrastructures requires cost-effective alternatives to conventional solid-core fiber (SCF). This paper examines using hollow-core fiber (HCF) with digital subcarrier multiplexing (DSCM) transceivers in horseshoe-and-spur filterless optical architectures. Leveraging HCF’s ultra-low nonlinearity, we optimize amplifier placement to maximize power budgets under realistic constraints. Our results show that HCF shifts the main limitation from nonlinearity to amplifier output power, enabling up to a 20 dB spur power-budget gain over SCF. Considering balanced and unbalanced couplers, we find that increasing amplifier density boosts reach only up to a saturation point (about 11–13 units in a 5-transit-node network). A techno-economic break-even analysis of hybrid SCF–HCF deployments shows that targeted HCF use provides intermediate performance gains and can fully recover its fiber cost premium through amplifier reductions.
\end{abstract}


\section{Introduction}\label{intro}

Historically, metro and access networks have been constrained by extreme cost-sensitive requirements. Hence, although advanced optical technologies were quickly adopted in long-haul segments and gradually became pervasive in regional and metropolitan segments, intensity modulation with direct detection (IM-DD) still dominates the edge. However, the maturity and volume production of coherent pluggable optics has drastically reduced the cost, power, and form-factor barriers to entry~\cite{pedro2025extended}. This shift facilitates a fundamental architectural transition: the convergence of metro and access layers into a unified, transparent optical infrastructure~\cite{cavaliere2025will}. By eliminating redundant optical-electrical-optical (O/E/O) regeneration at the network interface, operators can significantly simplify operations and reduce capital expenditures.

A key challenge in converged networks is the data-rate disparity across segments. Coherent point-to-multipoint (P2MP) transceivers address this by leveraging digital subcarrier multiplexing (DSCM). Unlike traditional point-to-point single-carrier systems, DSCM divides a high-capacity carrier into independent, lower-speed subcarriers (SCs)~\cite{welch2023digital}. This allows a single hub interface to simultaneously serve multiple distributed leaf nodes, making it highly effective for hub-and-spoke traffic. Consequently, DSCM provides fine-grained capacity allocation—such as configuring a 400\,Gbps envelope into $16 \times 25$\,Gbps or $4 \times 100$\,Gbps streams—while seamlessly maintaining compatibility with legacy passive optical networks (PONs)~\cite{simon2024200} and filterless architectures~\cite{hosseini2024optimized}. While the application of HCF is traditionally associated with long-haul transmission, its deployment in short-reach metro-access architectures—such as the horseshoe-and-spur topology—is increasingly motivated by the stringent requirements of emerging 5G and 6G Centralized-RAN (C-RAN) fronthaul networks. In these converged network environments, strict ultra-low latency budgets must be satisfied to meet critical timing constraints~\cite{shariati2021demonstration, nooruzzaman2024cost}. While DSCM optimizes the digital layer, hollow-core fiber (HCF) technology offers a parallel revolution at the transmission medium. By propagating light through a near-vacuum medium, HCF achieves a $50\%$ increase in propagation speed compared to traditional solid-core fibers (SCF), resulting in a $33\%$ latency reduction—a critical metric for high-speed trading and edge computing~\cite{borzycki2023hollow, petrovich2025broadband}. Furthermore, HCF exhibits negligible nonlinearity and stimulated Raman scattering (SRS), enabling higher launch powers and potentially extended reach~\cite{liu2018nonlinearity,correia2025hollow}. In metro-aggregation topologies, such as "horseshoe-and-spur" networks, the synergy of HCF’s low nonlinearity and DSCM’s subcarrier flexibility presents a unique opportunity to maximize scalability and power efficiency. Note that traditional fibers are limited by the Kerr nonlinearity, which creates a nonlinear Shannon limit~\cite{ellis2017performance}. HCF effectively moves this limit, allowing for much higher launch powers.

 \begin{figure}[h]
   \centering
 \centering\includegraphics[width=0.8\linewidth]{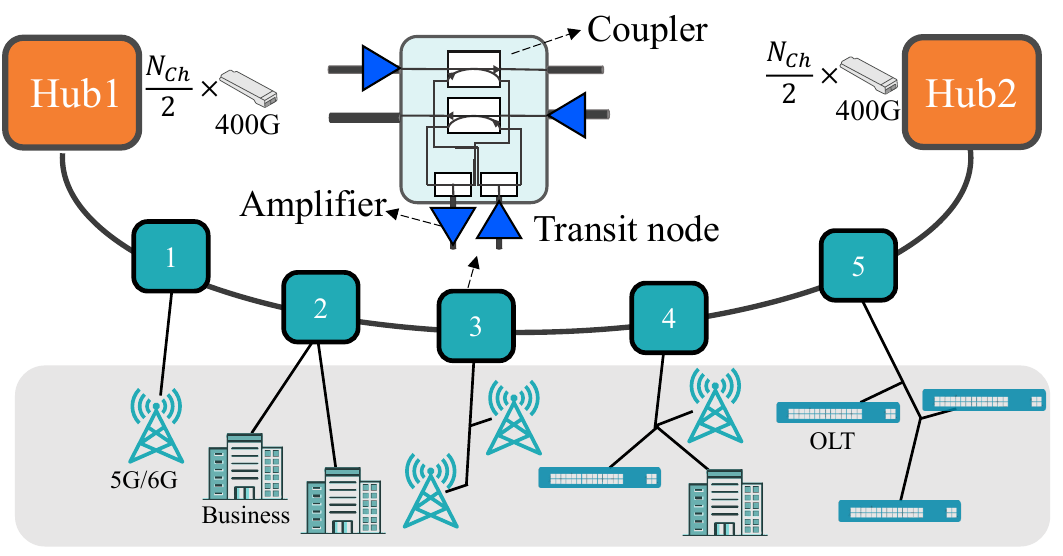}
 \caption{Horseshoe-and-spur optical network architecture.}
 \label{fig:topology}
 \end{figure}

Building upon prior research in filterless horseshoe-and-spur networks—where we optimized amplifier placement via integer linear programming (ILP) models~\cite{hosseini2024optimized} and analyzed high-power operation limits~\cite{hosseininonlinear} with SCF—this paper investigates the so far unexplored performance frontiers of HCF-based converged networks. 

While our previous work~\cite{hosseini2024optimized} established the foundational ILP framework for SCF networks, the integration of HCF introduces fundamentally different physical bottlenecks that require novel modeling considerations. Unlike traditional SCF models where channel power is strictly bounded by Kerr nonlinearity limits per span, replacing SCF with HCF shifts the primary system bottleneck from the transmission medium to the active equipment—specifically, the aggregate output power saturation of the optical amplifiers. Consequently, the methodological novelty of this extended framework lies not merely in relaxing nonlinearity parameters, but in redefining the optimization space for an amplifier-limited regime. This requires introducing new capacity-power coupling constraints (coupling the number of DSCM channels with the amplifier's total output power limit). By doing so, our model provides new system-level engineering insights, such as identifying the exact saturation point of diminishing returns in amplifier deployment. The contributions of this work are three-fold:

\begin{enumerate}

\item We define the number of optical amplifiers as a fixed resource budget and propose a framework to maximize the total system power budget.

\item We provide a comparative analysis of network scalability under full HCF deployment, traditional SCF limits, and a specifically defined Hybrid SCF-HCF scenario. To clarify our architectural approach, this hybrid deployment assumes the shared transit links of the main horseshoe ring utilize traditional SCF, while HCF is deployed exclusively on the dedicated access drop spurs. This strategic configuration leverages HCF’s high power-tolerance to overcome the severe insertion losses introduced by passive optical splitters at the branching nodes, allowing higher launch powers to reach the end-users without nonlinear penalties.

\item We provide a simplified method on how power budget can be consumed by different types of spurs.\end{enumerate}

The remainder of the paper is organized as follows. Section II describes the proposed network architecture and introduces the optimization framework utilized. The results of analyzing a set of horseshoe-and-spur network topologies are reported in Section III. Finally, Section IV presents the main conclusions of this work.

\section{Network Architecture and Optimization Framework}

The horseshoe topology is a robust and evolving network architecture, frequently utilized in metro-aggregation segments, primarily due to being well-suited to support the hub-and-spoke traffic pattern that prevails in these networks and the survivability features it provides. In this structure, communication is mainly concentrated between nodes and the exterior of the network, leading to the designation of one or two hub nodes that function as central aggregators and distributors, thus simplifying the architecture. Horseshoe topology inherently provides single-point failure protection by allowing all leaf nodes to maintain two separate paths to the two hub nodes, thus ensuring survivability to link or hub failures. The inclusion of spurs, that is, small, short trees branching off the main horseshoe, addresses long-term traffic growth and network expansion while efficiently using existing horseshoe leaf nodes as transit points. Within this extended design, i.e., horseshoe-and-spur architecture, the hubs and leaf nodes are defined as the origins and endpoints for all-optical flows, while the transit nodes handle the add, drop, and express functionalities using passive couplers, as illustrated in Fig.~\ref{fig:topology}. Optical amplifiers can be placed before the horseshoe transit nodes in both directions and at the spurs in both upstream and downstream directions~\cite{hosseini2024optimized}.

\subsection{Reach Model for Spurs}

As evaluated in \cite{hosseini2026design}, the network architecture is the primary factor dictating reach and scalability. While single-stage trees (using a centralized $1 \times N$ splitter) provide superior efficiency for scaling node density, multi-stage trees (using cascaded $50{:}50$ couplers) are valuable for extending reach across greater distances with minimal fiber deployment. 

Let $N$ represent the number of leaf nodes and $a$ represent the linear transmission factor per fiber segment. The required power budget ($P_b$) in dB to compensate for the worst-path power loss in each architecture is defined as follows:

\begin{equation}
    P_{b,\text{single}} = -10\log\!\left(\frac{a}{N}\right)
    \label{eq:Pb_single}
\end{equation}

\begin{equation}
    P_{b,\text{multi}} = -10\log\!\left[a\!\left(\frac{a}{2}\right)^{N-1}\right]
    \label{eq:Pb_multi}
\end{equation}

These equations demonstrate that single-stage trees experience a polynomial power penalty with respect to node count $N$, whereas multi-stage trees exhibit an exponential power penalty due to cascaded coupling losses. Consequently, this highlights the necessity of optimizing unbalanced coupler ratios for improved power distribution \cite{hosseini2026design}.

\subsection{HCF in Horseshoe-and-Spur Topology}

\begin{figure}[htb]
  \centering
\centering\includegraphics[width=0.85\linewidth]{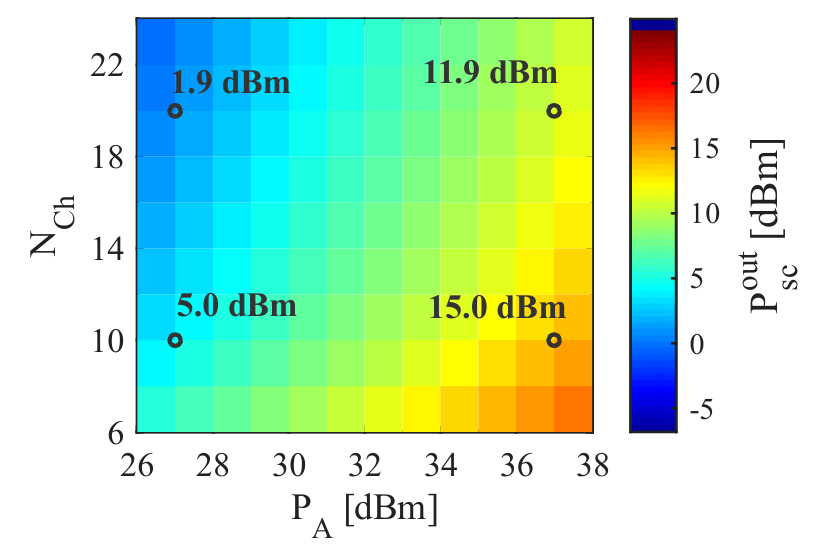}
\caption{SC power limit based on DSCM channel count and maximum amplifier output power.}
\label{fig:ampPower}
\end{figure}

If SCF within this topology is replaced by HCF fully or partly, several performance benefits and strategic considerations arise, particularly regarding transmission quality, latency, and cost. Although HCF is currently more expensive, its performance benefits can lead to cost reductions elsewhere in network planning and design. In addition, nonlinearity interference is substantially reduced in HCF because its nonlinear coefficient $\gamma$ is orders of magnitude lower than in SCF. In contrast to conventional SCF designs, which must honor limits to the launch power into the fiber to ensure a reduced impact of nonlinearities, with HCF, the amplifier's total output power can be the key limiting factor for increasing power levels. Consequently, for the case of an HCF fiber plant, the constraint $16\times N_{Ch}\times P_{SC}^{out} \leq P_{A}$ for 400G P2MP DSCM channels (with 16 SCs) can be added as additional constraints to our ILP framework described in~\cite{hosseini2024optimized}, while at the same time relaxing the nonlinearity constraints. Note that $P_{SC}^{out}$ denotes the output power per subcarrier, $N_{Ch}$ is the number of 400G DSCM channels, $P_{A}$ is the total output power of the optical amplifier, and the factor $16$ accounts for the number of subcarriers per 400G channel. Since there is dual-hub protection, $400G\times \frac{N_{Ch}}{2}$ will be the total capacity of the network, supporting a maximum of $8\times N_{Ch}$ end users. This constraint is visualized in Fig.~\ref{fig:ampPower}, which includes 4 selected points for the analysis in the following section.

The primary objective of our optimization framework is to maximize the total optical power budget across all access spurs ($\sum_{s=1}^{5}P_{b}^{s}$ [dB]). Maximizing this metric is essential because the available power budget dictates the maximum splitting ratio and fiber reach in passive branches, ensuring sufficient optical margin for future node expansions. For a fixed number of deployable amplifiers, this optimization is subject to three key constraints: (i) the received power at any endpoint must satisfy receiver sensitivity limits; (ii) the power imbalance among subcarriers must not exceed the receiver's tolerance; and (iii) the power budget allocation across spurs can be designated as either uniform or non-uniform.

% The optimization framework focuses on the maximization of the total power budgets of spurs (in logarithmic scale), i.e., $\sum_{s=1}^{5} P_b^s~[dB]$, facilitating future network expansion, as the new objective function, given a maximum number of amplifiers and subject to the constraints: (i) ensuring that the Rx input power matches or exceeds the receiver sensitivity, (ii) maintaining the SC power imbalance below the receiver tolerance (assuming a group of SCs coming from the same spur is already balanced), and (iii) enforcing (or not) uniformity in the spur power budget. 

\section{Results and Discussion}

We analyze 10 horseshoe-and-spur network architectures with five transit nodes and an average horseshoe link length of 12 km (see the distribution provided in~\cite{hosseini2024optimized}), utilizing couplers restricted to a balanced 50:50\% split or an unbalanced range from 50:50\% to 90:10\% in 10\% increments and assuming 0.5 dB excess loss. The internal configuration of the spurs is not explicitly predefined within this framework, allowing for any passive architecture—such as the single-stage or multi-stage trees evaluated in our analysis—provided it operates within the total available power budget. Rather than being restricted to a specific topology, the spurs are treated as flexible extensions where the primary constraint is the maximum power budget achieved through the strategic placement of amplifiers. This approach ensures that the optimized budget can accommodate various deployment scenarios, focusing on the worst-path power loss rather than a fixed physical layout. Our evaluation focused on two distinct power budget optimization scenarios for the spur connections: the Nonuniform power budget scenario, which maximizes the total power budget and allows spurs to receive varying individual allocations, and the Uniform power budget scenario, which enforces an explicit equality constraint requiring all spurs to have the same power budget. The parameters assumed in this study are summarized in Table~\ref{tab:network_parameters_compact}. We assume identical attenuation for HCF and SCF, although recent studies report record low-loss performance in HCF~\cite{petrovich2025broadband}.

\begin{table}[htb!]
    \centering
    \caption{Summary of network and transmission parameters.}
    \label{tab:network_parameters_compact}
    
    % 'p' allows text to wrap if the name is too long for the column width
    \begin{tabular}{p{0.55\columnwidth} l} 
        \hline
        \textbf{Parameter} & \textbf{Value} \\
        \hline
        Number of transit nodes & 5 \\
        \hline
        Number of channels & 10 and 20 \\
        \hline
        Modulation format & 16 QAM \\
        \hline
        Maximum EDFAs output power & $27$ dBm, $37$ dBm \\
        \hline
        Launch power & $-12$ dBm per SC \\
        \hline
        Sensitivity power at receiver & $-24$ dBm per SC \\
        \hline
        Nonlinear power threshold (SCF) & $-8$ dBm per SC \\
        \hline
        Nonlinear power threshold (HCF) & $10$ dBm per SC\\
        \hline
        Maximum SC power imbalance & $8$ dB \\
        \hline
        SCF loss & $0.24$ dB per km \\
        \hline
        HCF loss & $0.24$ dB per km \\
        \hline
        HCF splice loss & $0.2$ dB\\
        \hline
        Excess loss & $0.5$ dB\\
        \hline
        Average span length & $12$ km \\
        \hline
    \end{tabular}
\end{table}

\begin{figure*}[htb]
    \centering
    \scalebox{0.9}{ % <--- START scaling here
        \begin{minipage}{\linewidth} % 
            \subfloat[Balanced Couplers, Uniform Power Budget]{% This file was created by matlab2tikz.
\definecolor{teal_nokia}{rgb}{0.1372549019607843,0.6705882352941176,0.7137254901960784}%
\definecolor{blue_nokia}{rgb}{0 0.3529 1}%
\definecolor{orange_nokia}{rgb}{0.9568627450980393,0.498039,0.19215686274509805}%

\definecolor{mycolor2}{rgb}{0.12941,0.12941,0.12941}%
\begin{tikzpicture}[scale=1]

\begin{axis}[%
width=2.5in,
height=1.8in,
at={(0in,0in)},
scale only axis,
xmin=4,
xmax=16,
xlabel style={font=\color{mycolor2}},
xlabel={Number of Amplifiers},
ymin=0,
ymax=50,
ytick={0, 10, 20, 30, 40, 50},
ylabel style={font=\color{mycolor2}},
ylabel={Average Power Budget [dB]},
axis background/.style={fill=white},
xmajorgrids,
ymajorgrids,
legend columns=2,
legend style={legend cell align=left,
at={(0.994, 0.99)},
align=left,
font=\small}
]

\addplot[area legend, draw=none, fill=red, fill opacity=0.4, forget plot]
table[row sep=crcr] {%
x	y\\
6	2.32980873075957\\
7	4.7636358494694\\
8	10.4358838206247\\
9	13.3573326663406\\
10	16.7777056521322\\
11	20.1583592000051\\
12	22.5132558306529\\
13	25.1401954251788\\
14	26.4852386431896\\
15	27.9052720642064\\
16	27.974389730135\\
16	28.1881115147876\\
15	28.2177581528476\\
14	27.7447500608608\\
13	26.859906939524\\
12	24.0195466705137\\
11	21.9201608477596\\
10	20.0074006608715\\
9	15.6381817297279\\
8	14.9671207009358\\
7	7.66452147876109\\
6	3.89932207418807\\
}--cycle;
\addplot [color=red, line width=2.0pt, mark=*, mark options={solid, fill=red, red}]
  table[row sep=crcr]{%
6	3.11456540247382\\
7	6.21407866411525\\
8	12.7015022607803\\
9	14.4977571980342\\
10	18.3925531565018\\
11	21.0392600238823\\
12	23.2664012505833\\
13	26.0000511823514\\
14	27.1149943520252\\
15	28.061515108527\\
16	28.0812506224613\\
};
\addlegendentry{$\text{N}_{\text{Ch}}\text{ = 10, P}_{\text{A}}\text{=27}$}

\addplot[area legend, draw=none, fill=blue_nokia, fill opacity=0.4, forget plot]
table[row sep=crcr] {%
x	y\\
6	2.33227990661456\\
7	6.91359955290935\\
8	17.2650351191037\\
9	20.4707635164966\\
10	23.0898156584336\\
11	23.7403497182553\\
12	25.6881499409877\\
13	26.9362455466093\\
14	28.2794372747556\\
15	30.6826977154007\\
16	32.2029839759183\\
16	33.1475754694794\\
15	32.4869950010692\\
14	29.0850736664653\\
13	28.3194890820746\\
12	27.8402662861324\\
11	27.2235397322932\\
10	25.8982684964524\\
9	22.8495857866099\\
8	21.9359115245061\\
7	13.8478860077577\\
6	3.89953191198323\\
}--cycle;
\addplot [color=blue_nokia, line width=2.0pt, mark=*, mark options={solid, fill=blue_nokia, blue_nokia}]
  table[row sep=crcr]{%
6	3.11590590929889\\
7	10.3807427803335\\
8	19.6004733218049\\
9	21.6601746515532\\
10	24.494042077443\\
11	25.4819447252742\\
12	26.76420811356\\
13	27.627867314342\\
14	28.6822554706104\\
15	31.5848463582349\\
16	32.6752797226989\\
};
\addlegendentry{$\text{N}_{\text{Ch}}\text{ = 10, P}_{\text{A}}\text{=37}$}

\addplot[area legend, draw=none, fill=orange_nokia, fill opacity=0.4, forget plot]
table[row sep=crcr] {%
x	y\\
6	2.25779620382857\\
7	3.28474418788223\\
8	8.45164266362778\\
9	10.4206984923791\\
10	14.1866191258312\\
11	17.9545483260043\\
12	20.9099718467032\\
13	23.5065909995349\\
14	24.9819705417057\\
15	25.1504772277933\\
16	25.1277079256687\\
16	25.2542490177618\\
15	25.2579552693763\\
14	25.2914031910395\\
13	24.8145116151381\\
12	23.201117782743\\
11	19.7086515100549\\
10	17.3899744887293\\
9	12.6352712294205\\
8	12.0466954216588\\
7	5.69700403598818\\
6	3.81075329037358\\
}--cycle;
\addplot [color=orange_nokia, line width=2.0pt, mark=*, mark options={solid, fill=orange_nokia, orange_nokia}]
  table[row sep=crcr]{%
6	3.03427474710107\\
7	4.4908741119352\\
8	10.2491690426433\\
9	11.5279848608998\\
10	15.7882968072803\\
11	18.8315999180296\\
12	22.0555448147231\\
13	24.1605513073365\\
14	25.1366868663726\\
15	25.2042162485848\\
16	25.1909784717152\\
};
\addlegendentry{$\text{N}_{\text{Ch}}\text{ = 20, P}_{\text{A}}\text{=27}$}

\addplot[area legend, draw=none, fill=teal_nokia, fill opacity=0.4, forget plot]
table[row sep=crcr] {%
x	y\\
6	2.32904437874836\\
7	6.64480329654376\\
8	16.3253494519482\\
9	20.0301334082946\\
10	22.1992392449249\\
11	23.3477934097975\\
12	25.6284200130431\\
13	26.9417177694536\\
14	28.191809426017\\
15	30.0362064345945\\
16	32.1415221201646\\
16	33.0873529156982\\
15	32.4046870102528\\
14	28.9607924516225\\
13	28.2715598169402\\
12	27.7368564302871\\
11	26.8203429748911\\
10	25.4334868012703\\
9	22.0792191719022\\
8	21.1097860910965\\
7	12.810650507814\\
6	3.89333316609725\\
}--cycle;
\addplot [color=teal_nokia, line width=2.0pt, mark=*, mark options={solid, fill=teal_nokia, teal_nokia}]
  table[row sep=crcr]{%
6	3.11118877242281\\
7	9.7277269021789\\
8	18.7175677715224\\
9	21.0546762900984\\
10	23.8163630230976\\
11	25.0840681923443\\
12	26.6826382216651\\
13	27.6066387931969\\
14	28.5763009388197\\
15	31.2204467224236\\
16	32.6144375179314\\
};
\addlegendentry{$\text{N}_{\text{Ch}}\text{ = 20, P}_{\text{A}}\text{=37}$}

\addplot[area legend, draw=none, fill=black, fill opacity=0.4, forget plot]
table[row sep=crcr] {%
x	y\\
7	2.47649151753175\\
8	4.48983443415223\\
9	6.10439479522594\\
10	10.4448763357635\\
11	12.5180781749051\\
12	13.9961763975157\\
13	15.2134132600921\\
14	15.2781094324256\\
15	15.251399917131\\
16	15.2591340369748\\
16	15.4173647054709\\
15	15.4213283056742\\
14	15.4207174324296\\
13	15.3999377565821\\
12	15.3203906413491\\
11	14.2292501360801\\
10	11.7011037912746\\
9	9.35940099653382\\
8	6.31283028033945\\
7	3.91068052533928\\
}--cycle;
\addplot [color=black, dashed, line width=2.0pt, mark=*, mark options={solid, fill=black, black}]
  table[row sep=crcr]{%
7	3.19358602143551\\
8	5.40133235724584\\
9	7.73189789587988\\
10	11.0729900635191\\
11	13.3736641554926\\
12	14.6582835194324\\
13	15.3066755083371\\
14	15.3494134324276\\
15	15.3363641114026\\
16	15.3382493712228\\
};
\addlegendentry{Baseline: SCF}

\end{axis}

\end{tikzpicture}%
 \label{subfig:balanced_uniform}}%
            \hfill
            \subfloat[Balanced Couplers, Non-uniform Power Budget]{\definecolor{teal_nokia}{rgb}{0.1372549019607843,0.6705882352941176,0.7137254901960784}%
\definecolor{blue_nokia}{rgb}{0 0.3529 1}%
\definecolor{orange_nokia}{rgb}{0.9568627450980393,0.498039,0.19215686274509805}%
\definecolor{mycolor1}{rgb}{0.00000,1.00000,1.00000}%
\definecolor{mycolor2}{rgb}{0.12941,0.12941,0.12941}%
\begin{tikzpicture}[scale=1]

\begin{axis}[%
width=2.5in,
height=1.8in,
at={(0in,0in)},
scale only axis,
xmin=4,
xmax=16,
ytick={0, 10, 20, 30, 40, 50},
xlabel style={font=\color{mycolor2}},
xlabel={Number of Amplifiers},
ymin=0,
ymax=50,
ylabel style={font=\color{mycolor2}},
ylabel={Average Power Budget [dB]},
axis background/.style={fill=white},
xmajorgrids,
ymajorgrids,
% --- Add this line for a double-column legend ---
legend columns=2,
% ------------------------------------------------
legend style={legend cell align=left, 
at={(0.994, 0.99)},
align=left,
font=\small}
]

\addplot[area legend, draw=none, fill=red, fill opacity=0.4, forget plot]
table[row sep=crcr] {%
x	y\\
6	8.8792709\\
7	14.632852096\\
8	18.947570982\\
9	20.86978128\\
10	23.01202906\\
11	24.66668264\\
12	25.78672444\\
13	26.89500682\\
14	27.45364832\\
15	27.98378318\\
16	28.03315356\\
16	28.20088352\\
15	28.22260464\\
14	28.01155104\\
13	27.60107118\\
12	26.4759811\\
11	25.11222122\\
10	23.50473624\\
9	21.72695614\\
8	20.1318635\\
7	17.941874858\\
6	11.080566318\\
}--cycle;
\addplot [color=red, line width=2.0pt, mark=*, mark options={solid, fill=red, red}]
  table[row sep=crcr]{%
6	9.979918608\\
7	16.287363476\\
8	19.539717244\\
9	21.2983687\\
10	23.25838264\\
11	24.88945194\\
12	26.13135278\\
13	27.248039\\
14	27.73259968\\
15	28.1031939\\
16	28.11701854\\
};
\addlegendentry{$\text{N}_{\text{Ch}}\text{ = 10, P}_{\text{A}}\text{=27}$}

\addplot[area legend, draw=none, fill=blue_nokia, fill opacity=0.4, forget plot]
table[row sep=crcr] {%
x	y\\
6	11.384548048\\
7	17.119187034\\
8	22.31662376\\
9	25.76195282\\
10	28.05926376\\
11	29.5526033\\
12	30.27126132\\
13	30.9565106\\
14	31.3563926\\
15	32.10423292\\
16	32.7788789\\
16	33.18952284\\
15	32.88181474\\
14	32.1900658\\
13	31.6268238\\
12	31.05257448\\
11	30.51870988\\
10	29.47652136\\
9	27.57494564\\
8	24.0206123\\
7	20.48239514\\
6	14.845599198\\
}--cycle;
\addplot [color=blue_nokia, line width=2.0pt, mark=*, mark options={solid, fill=blue_nokia, blue_nokia}]
  table[row sep=crcr]{%
6	13.115073624\\
7	18.800791086\\
8	23.16861802\\
9	26.66844922\\
10	28.76789256\\
11	30.03565658\\
12	30.6619179\\
13	31.2916672\\
14	31.7732292\\
15	32.49302384\\
16	32.98420088\\
};
\addlegendentry{$\text{N}_{\text{Ch}}\text{ = 10, P}_{\text{A}}\text{=37}$}

\addplot[area legend, draw=none, fill=orange_nokia, fill opacity=0.4, forget plot]
table[row sep=crcr] {%
x	y\\
6	8.498296036\\
7	13.233574636\\
8	16.644341666\\
9	19.086630268\\
10	20.88167366\\
11	22.523465\\
12	23.7076703\\
13	24.62626186\\
14	24.98195182\\
15	25.01744648\\
16	25.10310954\\
16	25.27484348\\
15	25.13964838\\
14	25.1328761\\
13	25.05644054\\
12	24.44693432\\
11	23.33683738\\
10	21.65888082\\
9	19.84146097\\
8	18.159720194\\
7	15.704337766\\
6	10.200445306\\
}--cycle;
\addplot [color=orange_nokia, line width=2.0pt, mark=*, mark options={solid, fill=orange_nokia, orange_nokia}]
  table[row sep=crcr]{%
6	9.34937067\\
7	14.468956202\\
8	17.40203093\\
9	19.464045618\\
10	21.27027724\\
11	22.93015118\\
12	24.0773023\\
13	24.8413512\\
14	25.05741396\\
15	25.07854744\\
16	25.18897652\\
};
\addlegendentry{$\text{N}_{\text{Ch}}\text{ = 20, P}_{\text{A}}\text{=27}$}

\addplot[area legend, draw=none, fill=teal_nokia, fill opacity=0.4, forget plot]
table[row sep=crcr] {%
x	y\\
6	10.557753014\\
7	16.444896162\\
8	21.9605382\\
9	25.20141978\\
10	27.60127352\\
11	29.20365602\\
12	29.94301924\\
13	30.85778506\\
14	31.33996916\\
15	32.06110854\\
16	32.77465328\\
16	33.15038064\\
15	32.90931446\\
14	32.1212273\\
13	31.60708546\\
12	30.85828662\\
11	30.17270032\\
10	29.01267188\\
9	26.69862162\\
8	23.57980476\\
7	20.22572378\\
6	13.823148164\\
}--cycle;
\addplot [color=teal_nokia, line width=2.0pt, mark=*, mark options={solid, fill=teal_nokia, teal_nokia}]
  table[row sep=crcr]{%
6	12.19045059\\
7	18.335309976\\
8	22.77017148\\
9	25.9500207\\
10	28.3069727\\
11	29.68817818\\
12	30.40065294\\
13	31.23243526\\
14	31.73059824\\
15	32.4852115\\
16	32.96251696\\
};
\addlegendentry{$\text{N}_{\text{Ch}}\text{ = 20, P}_{\text{A}}\text{=37}$}

\addplot[area legend, draw=none, fill=black, fill opacity=0.4, forget plot]
table[row sep=crcr] {%
x	y\\
6	7.29405782001675\\
7	8.45657472508896\\
8	10.305741827436\\
9	12.4077377803507\\
10	13.6843568302702\\
11	14.5958230608975\\
12	15.0357228893669\\
13	15.2332451884742\\
14	15.2228569541999\\
15	15.2612025934485\\
16	15.2887813791309\\
16	15.3999146323641\\
15	15.4161177077269\\
14	15.4207526257309\\
13	15.3802236714206\\
12	15.3160895783645\\
11	14.9985383321619\\
10	14.1202228278337\\
9	13.1147085916536\\
8	11.1131437702474\\
7	9.53567014176483\\
6	7.90729209055224\\
}--cycle;
\addplot [color=black, dashed, line width=2.0pt, mark=*, mark options={solid, fill=black, black}]
  table[row sep=crcr]{%
6	7.60067495528449\\
7	8.99612243342689\\
8	10.7094427988417\\
9	12.7612231860022\\
10	13.902289829052\\
11	14.7971806965297\\
12	15.1759062338657\\
13	15.3067344299474\\
14	15.3218047899654\\
15	15.3386601505877\\
16	15.3443480057475\\
};
\addlegendentry{Baseline: SCF}

\end{axis}

\end{tikzpicture}%
 \label{subfig:balanced_nonuniform}}

            \subfloat[Unbalanced Couplers, Uniform Power Budget]{% This file was created by matlab2tikz.
\definecolor{teal_nokia}{rgb}{0.1372549019607843,0.6705882352941176,0.7137254901960784}%
\definecolor{blue_nokia}{rgb}{0 0.3529 1}%
\definecolor{orange_nokia}{rgb}{0.9568627450980393,0.498039,0.19215686274509805}%

\definecolor{mycolor2}{rgb}{0.12941,0.12941,0.12941}%
\begin{tikzpicture}[scale=1]

\begin{axis}[%
width=2.5in,
height=1.8in,
at={(0in,0in)},
scale only axis,
xmin=4,
xmax=16,
ytick={0, 10, 20, 30, 40, 50},
xlabel style={font=\color{mycolor2}},
xlabel={Number of Amplifiers},
ymin=0,
ymax=50,
ylabel style={font=\color{mycolor2}},
ylabel={Average Power Budget [dB]},
axis background/.style={fill=white},
xmajorgrids,
ymajorgrids,
legend columns=2,
legend style={legend cell align=left,
at={(0.994, 0.99)},
align=left,
font=\small}
]

\addplot[area legend, draw=none, fill=red, fill opacity=0.4, forget plot]
table[row sep=crcr] {%
x	y\\
4	5.21455753269161\\
5	7.39664442803256\\
6	12.91548351332\\
7	17.1067808015621\\
8	19.4841499801986\\
9	23.5937040942225\\
10	24.3127585556531\\
11	26.9738376826086\\
12	28.5977000533713\\
13	28.676282544743\\
14	28.6811960032351\\
15	28.682282001375\\
16	28.6803377004816\\
16	28.6949632419303\\
15	28.7046081293534\\
14	28.7007019684665\\
13	28.7017778158093\\
12	28.6835846041983\\
11	28.500915470127\\
10	25.1582300094639\\
9	24.3799651521655\\
8	21.5248692576498\\
7	18.571663409759\\
6	15.9596923091744\\
5	10.1716434014095\\
4	7.54699620883473\\
}--cycle;
\addplot [color=red, line width=2.0pt, mark=*, mark options={solid, fill=red, red}]
  table[row sep=crcr]{%
4	6.38077687076317\\
5	8.78414391472103\\
6	14.4375879112472\\
7	17.8392221056606\\
8	20.5045096189242\\
9	23.986834623194\\
10	24.7354942825585\\
11	27.7373765763678\\
12	28.6406423287848\\
13	28.6890301802761\\
14	28.6909489858508\\
15	28.6934450653642\\
16	28.6876504712059\\
};
\addlegendentry{$\text{N}_{\text{Ch}}\text{ = 10, P}_{\text{A}}\text{=27}$}

\addplot[area legend, draw=none, fill=blue_nokia, fill opacity=0.4, forget plot]
table[row sep=crcr] {%
x	y\\
4	5.9291923814191\\
5	8.59234362521461\\
6	18.6224753057361\\
7	22.0287194115853\\
8	24.6535297643009\\
9	27.4667778743214\\
10	29.3444851355323\\
11	31.3858920537102\\
12	32.0670498518891\\
13	33.1660532827834\\
14	33.6652053464898\\
15	33.6783441541116\\
16	33.6714130011667\\
16	33.7029773212314\\
15	33.7069209677937\\
14	33.7132036969823\\
13	33.7001345216185\\
12	32.9734834227746\\
11	32.3365995901097\\
10	31.6354104147113\\
9	30.2336926477685\\
8	26.5233252522819\\
7	23.7711501904044\\
6	21.9258776305365\\
5	12.6454049632443\\
4	10.5007875877321\\
}--cycle;
\addplot [color=blue_nokia, line width=2.0pt, mark=*, mark options={solid, fill=blue_nokia, blue_nokia}]
  table[row sep=crcr]{%
4	8.21498998457562\\
5	10.6188742942295\\
6	20.2741764681363\\
7	22.8999348009948\\
8	25.5884275082914\\
9	28.850235261045\\
10	30.4899477751218\\
11	31.8612458219099\\
12	32.5202666373319\\
13	33.4330939022009\\
14	33.6892045217361\\
15	33.6926325609526\\
16	33.6871951611991\\
};
\addlegendentry{$\text{N}_{\text{Ch}}\text{ = 10, P}_{\text{A}}\text{=37}$}

\addplot[area legend, draw=none, fill=orange_nokia, fill opacity=0.4, forget plot]
table[row sep=crcr] {%
x	y\\
4	4.77173664096401\\
5	6.23718962015669\\
6	10.5699856851469\\
7	14.4832229531972\\
8	18.0638407718577\\
9	21.1143811391329\\
10	22.5611380699873\\
11	25.2530818374475\\
12	25.6932658482796\\
13	25.6932457558442\\
14	25.6932995734837\\
15	25.7050375953077\\
16	25.6948570927444\\
16	25.7132746074025\\
15	25.7347733705751\\
14	25.7075974459623\\
13	25.7254801142713\\
12	25.7180570370437\\
11	25.8037049040206\\
10	24.430540359632\\
9	22.8183406563855\\
8	20.4327675397432\\
7	16.1537178420588\\
6	13.3023172408419\\
5	8.58489894534931\\
4	6.98298806727601\\
}--cycle;
\addplot [color=orange_nokia, line width=2.0pt, mark=*, mark options={solid, fill=orange_nokia, orange_nokia}]
  table[row sep=crcr]{%
4	5.87736235412001\\
5	7.41104428275299\\
6	11.9361514629944\\
7	15.318470397628\\
8	19.2483041558005\\
9	21.9663608977592\\
10	23.4958392148097\\
11	25.5283933707341\\
12	25.7056614426617\\
13	25.7093629350578\\
14	25.700448509723\\
15	25.7199054829414\\
16	25.7040658500735\\
};
\addlegendentry{$\text{N}_{\text{Ch}}\text{ = 20, P}_{\text{A}}\text{=27}$}

\addplot[area legend, draw=none, fill=teal_nokia, fill opacity=0.4, forget plot]
table[row sep=crcr] {%
x	y\\
4	5.8924912237877\\
5	8.5831337624444\\
6	17.692304815275\\
7	21.3135308845572\\
8	23.6038385587207\\
9	26.8499755307683\\
10	28.6672154196152\\
11	30.4247389099488\\
12	31.5631346018835\\
13	32.9687108969162\\
14	33.4433790829404\\
15	33.6796382941393\\
16	33.6811996068542\\
16	33.710482639489\\
15	33.7234015646347\\
14	33.7297345158365\\
13	33.7403206868018\\
12	32.7959557646459\\
11	31.4752216476917\\
10	30.4724520288995\\
9	29.4678568588738\\
8	25.4145874812852\\
7	22.576438990433\\
6	20.8511855674393\\
5	12.5683449148425\\
4	10.3269939445136\\
}--cycle;
\addplot [color=teal_nokia, line width=2.0pt, mark=*, mark options={solid, fill=teal_nokia, teal_nokia}]
  table[row sep=crcr]{%
4	8.10974258415065\\
5	10.5757393386434\\
6	19.2717451913571\\
7	21.9449849374951\\
8	24.509213020003\\
9	28.158916194821\\
10	29.5698337242574\\
11	30.9499802788202\\
12	32.1795451832647\\
13	33.354515791859\\
14	33.5865567993884\\
15	33.701519929387\\
16	33.6958411231716\\
};
\addlegendentry{$\text{N}_{\text{Ch}}\text{ = 20, P}_{\text{A}}\text{=37}$}

\addplot[area legend, draw=none, fill=black, fill opacity=0.4, forget plot]
table[row sep=crcr] {%
x	y\\
5	5.95017478587698\\
6	8.07358162767092\\
7	10.2930389752778\\
8	15.5938645846448\\
9	15.8051989326921\\
10	15.7937857193771\\
11	15.7978834014109\\
12	15.8245294871921\\
13	15.8120447211712\\
14	15.8336518506146\\
15	15.8234696988347\\
16	15.8309845682957\\
16	15.8493610253165\\
15	15.8629889052\\
14	15.8548242154228\\
13	15.847084195927\\
12	15.8492268702899\\
11	15.8485967171316\\
10	15.8433445228401\\
9	15.8353502529051\\
8	15.8293128365065\\
7	12.336109201357\\
6	9.05708907302273\\
5	7.59131003051611\\
}--cycle;
\addplot [color=black, dashed, line width=2.0pt, mark=*, mark options={solid, fill=black, black}]
  table[row sep=crcr]{%
5	6.77074240819655\\
6	8.56533535034683\\
7	11.3145740883174\\
8	15.7115887105756\\
9	15.8202745927986\\
10	15.8185651211086\\
11	15.8232400592712\\
12	15.836878178741\\
13	15.8295644585491\\
14	15.8442380330187\\
15	15.8432293020173\\
16	15.8401727968061\\
};
\addlegendentry{Baseline: SCF}

\end{axis}

\end{tikzpicture}%
 \label{subfig:unbalanced_uniform}}%
            \hfill
            \subfloat[Unbalanced Couplers, Non-uniform Power Budget]{% This file was created by matlab2tikz.
\definecolor{teal_nokia}{rgb}{0.1372549019607843,0.6705882352941176,0.7137254901960784}%
\definecolor{blue_nokia}{rgb}{0 0.3529 1}%
\definecolor{orange_nokia}{rgb}{0.9568627450980393,0.498039,0.19215686274509805}%

\definecolor{mycolor2}{rgb}{0.12941,0.12941,0.12941}%
\begin{tikzpicture}[scale=1]

\begin{axis}[%
width=2.5in,
height=1.8in,
at={(0in,0in)},
scale only axis,
xmin=4,
xmax=16,
ytick={0, 10, 20, 30, 40, 50},
xlabel style={font=\color{mycolor2}},
xlabel={Number of Amplifiers},
ymin=0,
ymax=50,
ylabel style={font=\color{mycolor2}},
ylabel={Average Power Budget [dB]},
axis background/.style={fill=white},
xmajorgrids,
ymajorgrids,
legend columns=2,
legend style={legend cell align=left,
at={(0.994, 0.99)},
align=left,
font=\small}
]

\addplot[area legend, draw=none, fill=red, fill opacity=0.4, forget plot]
table[row sep=crcr] {%
x	y\\
4	6.32076940426716\\
5	10.9832759339603\\
6	15.3853724777477\\
7	19.9000409420964\\
8	23.234213529745\\
9	26.0737017262039\\
10	27.1005215951472\\
11	27.9525804511717\\
12	28.615157611289\\
13	28.6791440176227\\
14	28.6843834966353\\
15	28.682095983639\\
16	28.6797423766527\\
16	28.7107769113095\\
15	28.7111614309912\\
14	28.724502645835\\
13	28.7015564535248\\
12	28.6952676279484\\
11	28.5544428359343\\
10	27.6176374958204\\
9	26.5477162604661\\
8	24.1212999851847\\
7	21.96816238845\\
6	18.7944620035636\\
5	13.7155687007399\\
4	8.5424754470115\\
}--cycle;
\addplot [color=red, line width=2.0pt, mark=*, mark options={solid, fill=red, red}]
  table[row sep=crcr]{%
4	7.43162242563933\\
5	12.3494223173501\\
6	17.0899172406556\\
7	20.9341016652732\\
8	23.6777567574649\\
9	26.310708993335\\
10	27.3590795454838\\
11	28.253511643553\\
12	28.6552126196187\\
13	28.6903502355737\\
14	28.7044430712351\\
15	28.6966287073151\\
16	28.6952596439811\\
};
\addlegendentry{$\text{N}_{\text{Ch}}\text{ = 10, P}_{\text{A}}\text{=27}$}

\addplot[area legend, draw=none, fill=blue_nokia, fill opacity=0.4, forget plot]
table[row sep=crcr] {%
x	y\\
4	7.1826367531394\\
5	11.7880164987109\\
6	19.9270362788295\\
7	24.6224627330493\\
8	27.4400913482534\\
9	30.2357882090546\\
10	31.898783991077\\
11	32.6595869015007\\
12	33.1313219341014\\
13	33.4830014649115\\
14	33.6743050641762\\
15	33.675044859715\\
16	33.6601420873436\\
16	33.7054584432457\\
15	33.7233383180571\\
14	33.6955178132131\\
13	33.6991514877507\\
12	33.4002512657349\\
11	33.1247163369705\\
10	32.683816609615\\
9	31.8633191276258\\
8	29.5212574685471\\
7	26.2826717147518\\
6	22.4568125223238\\
5	15.6709263165091\\
4	11.3644070188249\\
}--cycle;
\addplot [color=blue_nokia, line width=2.0pt, mark=*, mark options={solid, fill=blue_nokia, blue_nokia}]
  table[row sep=crcr]{%
4	9.27352188598213\\
5	13.72947140761\\
6	21.1919244005767\\
7	25.4525672239005\\
8	28.4806744084002\\
9	31.0495536683402\\
10	32.291300300346\\
11	32.8921516192356\\
12	33.2657865999182\\
13	33.5910764763311\\
14	33.6849114386946\\
15	33.6991915888861\\
16	33.6828002652947\\
};
\addlegendentry{$\text{N}_{\text{Ch}}\text{ = 10, P}_{\text{A}}\text{=37}$}

\addplot[area legend, draw=none, fill=orange_nokia, fill opacity=0.4, forget plot]
table[row sep=crcr] {%
x	y\\
4	5.71465899805562\\
5	10.1431459173002\\
6	14.5253328607303\\
7	18.2955799983056\\
8	21.270674915712\\
9	23.4976452422663\\
10	24.6441915518212\\
11	25.5016474181981\\
12	25.6969647133838\\
13	25.6933849797241\\
14	25.7024434366885\\
15	25.6991325724458\\
16	25.6970567050576\\
16	25.7242129445598\\
15	25.7213638587645\\
14	25.7286383878998\\
13	25.7004167700529\\
12	25.7208630562265\\
11	25.7571089125753\\
10	25.0828474821433\\
9	24.1013106046149\\
8	22.9168039893321\\
7	20.6180446784014\\
6	17.0180748138927\\
5	12.1835922989204\\
4	7.82470167340528\\
}--cycle;
\addplot [color=orange_nokia, line width=2.0pt, mark=*, mark options={solid, fill=orange_nokia, orange_nokia}]
  table[row sep=crcr]{%
4	6.76968033573045\\
5	11.1633691081103\\
6	15.7717038373115\\
7	19.4568123383535\\
8	22.0937394525221\\
9	23.7994779234406\\
10	24.8635195169822\\
11	25.6293781653867\\
12	25.7089138848051\\
13	25.6969008748885\\
14	25.7155409122942\\
15	25.7102482156051\\
16	25.7106348248087\\
};
\addlegendentry{$\text{N}_{\text{Ch}}\text{ = 20, P}_{\text{A}}\text{=27}$}

\addplot[area legend, draw=none, fill=teal_nokia, fill opacity=0.4, forget plot]
table[row sep=crcr] {%
x	y\\
4	7.08732313474811\\
5	11.7751799612319\\
6	19.2449611867791\\
7	24.2234578525427\\
8	27.1812936826213\\
9	29.7364183804967\\
10	31.2644215148119\\
11	32.4064013671501\\
12	32.8068017051863\\
13	33.3727437950437\\
14	33.6050315537791\\
15	33.6702493998379\\
16	33.666392232261\\
16	33.6957558713541\\
15	33.7019572919652\\
14	33.7155329750691\\
13	33.685303479196\\
12	33.2972971826851\\
11	32.8692955352145\\
10	32.03918544271\\
9	31.491706144805\\
8	28.9850892484444\\
7	25.532194797517\\
6	21.4682963341476\\
5	15.6292636730471\\
4	11.1173193567387\\
}--cycle;
\addplot [color=teal_nokia, line width=2.0pt, mark=*, mark options={solid, fill=teal_nokia, teal_nokia}]
  table[row sep=crcr]{%
4	9.10232124574342\\
5	13.7022218171395\\
6	20.3566287604634\\
7	24.8778263250298\\
8	28.0831914655328\\
9	30.6140622626508\\
10	31.651803478761\\
11	32.6378484511823\\
12	33.0520494439357\\
13	33.5290236371198\\
14	33.6602822644241\\
15	33.6861033459016\\
16	33.6810740518075\\
};
\addlegendentry{$\text{N}_{\text{Ch}}\text{ = 20, P}_{\text{A}}\text{=37}$}

\addplot[area legend, draw=none, fill=black, fill opacity=0.4, forget plot]
table[row sep=crcr] {%
x	y\\
5	8.13754743425618\\
6	11.2059807932073\\
7	13.4883358816952\\
8	15.6310050029529\\
9	15.760485572208\\
10	15.781165645529\\
11	15.798728951437\\
12	15.8059505944315\\
13	15.8302065479844\\
14	15.8237250767511\\
15	15.8309732411021\\
16	15.8368862334367\\
16	15.8515237311679\\
15	15.8512229851669\\
14	15.8509861669331\\
13	15.8498308080572\\
12	15.8483943225809\\
11	15.8516463526624\\
10	15.8299311867626\\
9	15.8198516891204\\
8	15.8080925522777\\
7	14.0861407439421\\
6	11.9923051634409\\
5	9.92004610594692\\
}--cycle;
\addplot [color=black, dashed, line width=2.0pt, mark=*, mark options={solid, fill=black, black}]
  table[row sep=crcr]{%
5	9.02879677010155\\
6	11.5991429783241\\
7	13.7872383128187\\
8	15.7195487776153\\
9	15.7901686306642\\
10	15.8055484161458\\
11	15.8251876520497\\
12	15.8271724585062\\
13	15.8400186780208\\
14	15.8373556218421\\
15	15.8410981131345\\
16	15.8442049823023\\
};
\addlegendentry{Baseline: SCF}

\end{axis}

\end{tikzpicture}%
 \label{subfig:unbalanced_nonuniform}}
        \end{minipage} % End minipage
    } % <--- END scaling here
    
    \caption{Comparison of average power budget per spur versus the number of amplifiers. The subfigures show the results for different combinations of coupler types  and power budget distribution.}    
   \label{fig:combined_results}
\end{figure*}
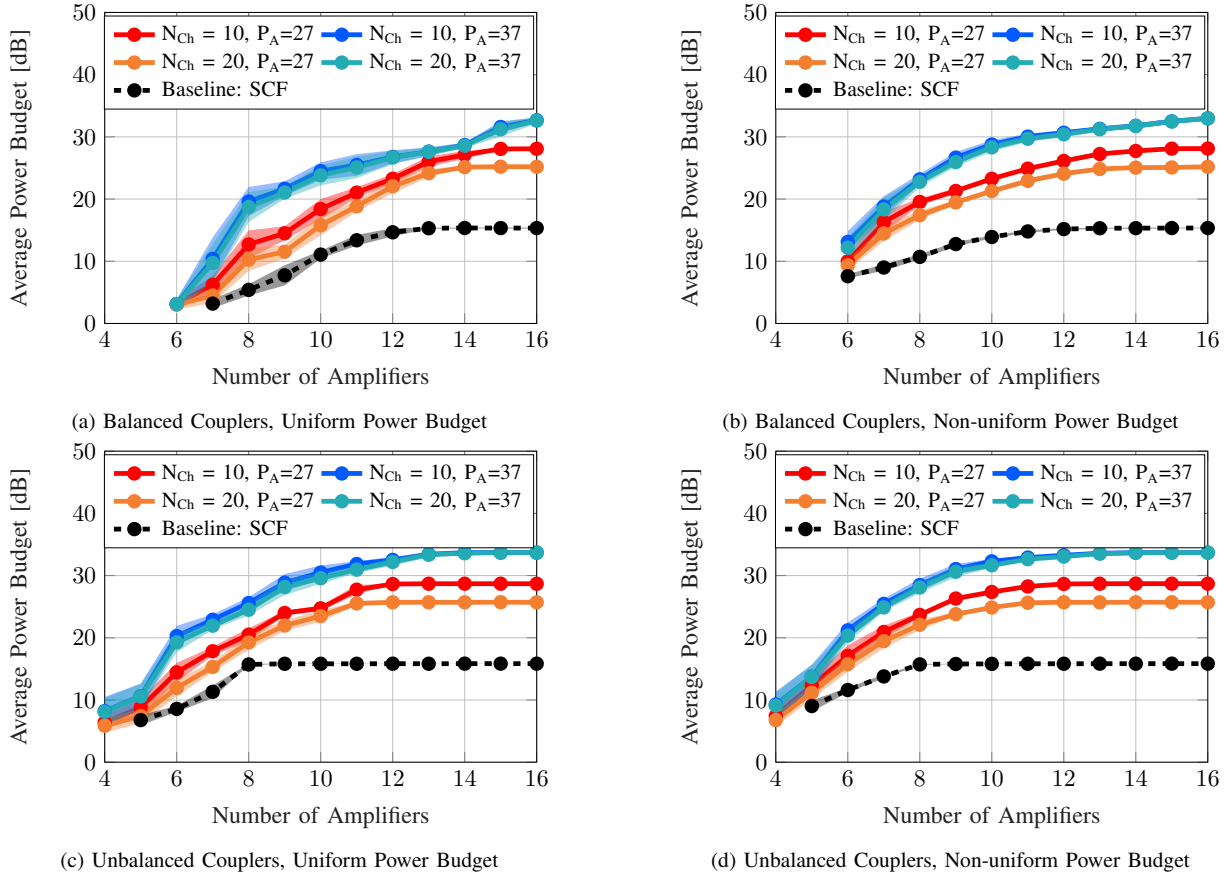
\vspace{-2mm}

Figures~\ref{fig:combined_results}(a) and (b) present a comprehensive comparison of the average power budget per spur as a function of the number of amplifiers that can be deployed for the four different cases of amplifier maximum total output power and number of channels highlighted in Fig.~\ref{fig:ampPower}. The plots also include the 90\% confidence intervals of the average results of the 10 networks considered. Importantly, these results demonstrate the substantial performance advantage of the HCF-based network over the SCF baseline. In both uniform and nonuniform power budget scenarios, HCF provides a significant increase in power budget, especially for amplifier counts greater than eight. This performance gain correlates positively with higher amplifier output power ($P_A$) and inversely with the number of channels ($P_{Ch}$). While the nonuniform scenario initially shows an advantage for fewer amplifiers, both scenarios eventually converge to similar power levels for a sufficiently high amplifier count; this saturation occurs because the nonlinearity power threshold (in the case of the SCF network) and the maximum output power of the amplifiers (in the case of the HCF network) become the limiting factors. This saturation effect is critical, as it shows that the total number of possible amplifiers (20 in this case) is not necessary to maximize the power budget, highlighting the importance of selective amplifier deployment. For a design employing unbalanced coupler ratios, Fig.~\ref{fig:combined_results}(c) and (d) demonstrate that a minimum of four amplifiers is essential to satisfy the design constraints, yielding an average power budget of only approximately 10 dB. With more amplifiers deployed, the SCF network reaches a maximum of 15.8 dB, whereas the HCF network provides a considerably higher power budget, ranging from 26 dB to 35 dB. Crucially, the unbalanced coupler ratio configuration offers a superior power budget compared to setups using balanced ratios. The diminishing returns in power budget gains occur because the network's average link length of 12 km and low fiber attenuation (0.24 dB/km) mean that not every potential node requires an amplifier to maintain signal quality. As a result, the system reaches a saturation point—in this case between 11 and 13 units—where adding more amplifiers no longer significantly improves performance. Beyond this point, the network's scalability is no longer limited by fiber loss, but rather by physical constraints such as nonlinearity or maximum output power of the amplifiers themselves. In our numerical evaluation, we assume an identical attenuation coefficient of 0.24 dB/km for both SCF and HCF. While recent state-of-the-art laboratory demonstrations have reported HCF attenuation below 0.1 dB/km~\cite{petrovich2025broadband}, our assumption of 0.24 dB/km serves two specific purposes. First, as a deliberate modeling simplification, it isolates the performance improvements derived solely from the absence of Kerr nonlinearities in HCF, allowing for a strictly fair comparison of power-budget gains uncoupled from propagation loss advantages. Second, it provides a conservative estimate for early-stage HCF types~\cite{jasion2020hollow}, where excess losses from field cabling, multiple splices, and bending are expected to increase the effective span attenuation compared to pristine laboratory conditions.

It is important to note that while our ILP framework primarily optimizes the power budget while considering nonlinear threshold constraints, moving to an amplifier-limited regime necessitates careful consideration of Amplified Spontaneous Emission (ASE) noise. Assuming a standard Erbium-Doped Fiber Amplifier (EDFA), the accumulated ASE noise in our targeted short-reach metro-access networks remains sufficiently low as we showed in previous studies~\cite{hosseini2026design, pedro2025extended}.

In certain brownfield deployments where horseshoe network architectures are already established, a hybrid horseshoe-and-spur model can be implemented by upgrading the spurs to HCF. This configuration significantly mitigates nonlinear impairments within the spur segments; however, the original SCF horseshoe remains power-limited. Consequently, the overall system performance offers a middle ground between a baseline SCF network and a full HCF system.

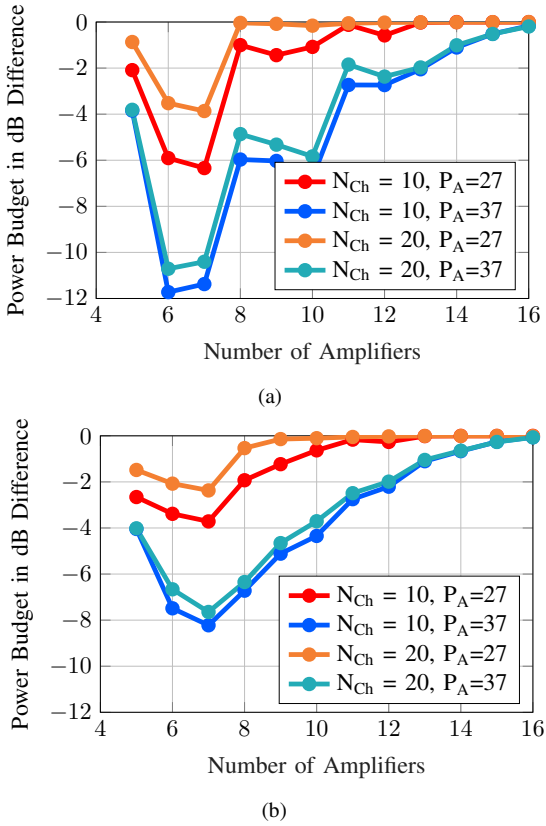
\begin{figure}[htb]
    \centering
    \scalebox{0.9}{ % <--- START scaling here
        \begin{minipage}{\linewidth} % 
            \subfloat[]{% This file was created by matlab2tikz.
\definecolor{teal_nokia}{rgb}{0.1372549019607843,0.6705882352941176,0.7137254901960784}%
\definecolor{blue_nokia}{rgb}{0 0.3529 1}%
\definecolor{orange_nokia}{rgb}{0.9568627450980393,0.498039,0.19215686274509805}%

\definecolor{mycolor1}{rgb}{0.00000,1.00000,1.00000}%
\definecolor{mycolor2}{rgb}{0.12941,0.12941,0.12941}%
\begin{tikzpicture}

\begin{axis}[%
width=2.5in,
height=1.6in,
at={(0.0in,0.0in)},
scale only axis,
xmin=4,
xmax=16,
ymin=-12,
ymax=0,
xlabel style={font=\color{mycolor2}},
xlabel={Number of Amplifiers},
ylabel style={font=\color{mycolor2}},
ylabel={Power Budget in dB Difference},
ylabel near ticks,
ytick={-12, -10, -8, -6, -4, -2, 0},
axis background/.style={fill=white},
xmajorgrids,
ymajorgrids,
legend style={legend cell align=left, align=left},
legend pos=south east
]
\addplot [color=red, line width=2.0pt, mark=*, mark options={solid, fill=red, red}]
  table[row sep=crcr]{%
5	-2.08867527846674\\
6	-5.90657258185898\\
7	-6.3397198439754\\
8	-1.00317199964236\\
9	-1.43882369140002\\
10	-1.08022717818375\\
11	-0.121582787768773\\
12	-0.581606652712161\\
13	-0.0228363095268485\\
14	-0.00265150112849355\\
15	-0.00720872029954833\\
16	-0.00485572443522031\\
};
\addlegendentry{$\text{N}_{\text{Ch}}\text{ = 10, P}_{\text{A}}\text{=27}$}

\addplot [color=blue_nokia, line width=2.0pt, mark=*, mark options={solid, fill=blue_nokia, blue_nokia}]
  table[row sep=crcr]{%
5	-3.84451946054763\\
6	-11.7245770967241\\
7	-11.3745121181425\\
8	-5.9690246829788\\
9	-6.02953010711785\\
10	-6.76821427636159\\
11	-2.73280448758156\\
12	-2.73734352327552\\
13	-2.05130014973469\\
14	-1.1057238920507\\
15	-0.515005506484023\\
16	-0.167502060347708\\
};
\addlegendentry{$\text{N}_{\text{Ch}}\text{ = 10, P}_{\text{A}}\text{=37}$}

\addplot [color=orange_nokia, line width=2.0pt, mark=*, mark options={solid, fill=orange_nokia, orange_nokia}]
  table[row sep=crcr]{%
5	-0.868262891244144\\
6	-3.52222048782998\\
7	-3.86396133130935\\
8	-0.0377914086147868\\
9	-0.0788380372939521\\
10	-0.158843873540686\\
11	-0.0551015233301051\\
12	-0.0254637592705826\\
13	-0.0103798100786605\\
14	-0.00933673754306596\\
15	-0.00420961679567\\
16	-0.000548406872592722\\
};
\addlegendentry{$\text{N}_{\text{Ch}}\text{ = 20, P}_{\text{A}}\text{=27}$}

\addplot [color=teal_nokia, line width=2.0pt, mark=*, mark options={solid, fill=teal_nokia, teal_nokia}]
  table[row sep=crcr]{%
5	-3.81107340025239\\
6	-10.7096410077906\\
7	-10.4056165831527\\
8	-4.86567180501373\\
9	-5.32295867296267\\
10	-5.83402010967978\\
11	-1.847569822755\\
12	-2.36944244968288\\
13	-1.97286940747374\\
14	-1.01072023606781\\
15	-0.520126266901208\\
16	-0.194508914258002\\
};
\addlegendentry{$\text{N}_{\text{Ch}}\text{ = 20, P}_{\text{A}}\text{=37}$}

\end{axis}

\end{tikzpicture}%
}%
            \hfill
            \subfloat[]{% This file was created by matlab2tikz.
\definecolor{teal_nokia}{rgb}{0.1372549019607843,0.6705882352941176,0.7137254901960784}%
\definecolor{blue_nokia}{rgb}{0 0.3529 1}%
\definecolor{orange_nokia}{rgb}{0.9568627450980393,0.498039,0.19215686274509805}%

\definecolor{mycolor1}{rgb}{0.00000,1.00000,1.00000}%
\definecolor{mycolor2}{rgb}{0.12941,0.12941,0.12941}%
\begin{tikzpicture}

\begin{axis}[%
width=2.5in,
height=1.6in,
at={(0.0in,0.0in)},
scale only axis,
xmin=4,
xmax=16,
ymin=-12,
ymax=0,
xlabel style={font=\color{mycolor2}},
xlabel={Number of Amplifiers},
ylabel style={font=\color{mycolor2}},
ylabel={Power Budget in dB Difference},
ylabel near ticks,
ytick={-12, -10, -8, -6, -4, -2, 0},
axis background/.style={fill=white},
xmajorgrids,
ymajorgrids,
legend style={legend cell align=left, align=left},
legend pos=south east
]
\addplot [color=red, line width=2.0pt, mark=*, mark options={solid, fill=red, red}]
  table[row sep=crcr]{%
5	-2.65719533121201\\
6	-3.38576446041538\\
7	-3.71495985763125\\
8	-1.92835971810563\\
9	-1.23108045458524\\
10	-0.6309501824819\\
11	-0.165533143590167\\
12	-0.265627683740036\\
13	-0.0124172172065435\\
14	-0.00315291167890308\\
15	-0.00520708925198932\\
16	-0.00274428499637835\\
};
\addlegendentry{$\text{N}_{\text{Ch}}\text{ = 10, P}_{\text{A}}\text{=27}$}

\addplot [color=blue_nokia, line width=2.0pt, mark=*, mark options={solid, fill=blue_nokia, blue_nokia}]
  table[row sep=crcr]{%
5	-4.04300058637606\\
6	-7.48667536403089\\
7	-8.22376392420417\\
8	-6.72078200976787\\
9	-5.11475823998186\\
10	-4.34566749437139\\
11	-2.74923295635033\\
12	-2.20822070879825\\
13	-1.10263224855701\\
14	-0.66563079207598\\
15	-0.258034527020669\\
16	-0.0587986203220012\\
};
\addlegendentry{$\text{N}_{\text{Ch}}\text{ = 10, P}_{\text{A}}\text{=37}$}

\addplot [color=orange_nokia, line width=2.0pt, mark=*, mark options={solid, fill=orange_nokia, orange_nokia}]
  table[row sep=crcr]{%
5	-1.48180478172061\\
6	-2.07568057740424\\
7	-2.37034129254882\\
8	-0.533331452136963\\
9	-0.141882455089309\\
10	-0.102984828923894\\
11	-0.0494471076715826\\
12	-0.0244602813315566\\
13	-0.00519276339477059\\
14	-0.00590181232525211\\
15	-0.00346143568244983\\
16	-0.0114139773173001\\
};
\addlegendentry{$\text{N}_{\text{Ch}}\text{ = 20, P}_{\text{A}}\text{=27}$}

\addplot [color=teal_nokia, line width=2.0pt, mark=*, mark options={solid, fill=teal_nokia, teal_nokia}]
  table[row sep=crcr]{%
5	-4.01570974528285\\
6	-6.66172611627034\\
7	-7.64564755478283\\
8	-6.34605059974618\\
9	-4.6565312993251\\
10	-3.70788596846796\\
11	-2.49005415877159\\
12	-1.98867688461969\\
13	-1.04690263141611\\
14	-0.645174924026662\\
15	-0.266817852429526\\
16	-0.0683228425504083\\
};
\addlegendentry{$\text{N}_{\text{Ch}}\text{ = 20, P}_{\text{A}}\text{=37}$}

\end{axis}

\end{tikzpicture}%
 }
        \end{minipage} % End minipage
    } % <--- END scaling here
    \caption{Power budget difference between HCF and SCF-HCF scenario with unbalanced couplers and (a) uniform and (b) non-uniform power budget.}    
   \label{fig:Diff_results}
\end{figure}

We evaluated the proposed hybrid architecture using unbalanced couplers, performing a comparative analysis of the power budget between pure HCF and mixed SCF-HCF configuration. Fig. \ref{fig:Diff_results} illustrates the performance gap in power budget between the hybrid SCF-HCF and pure HCF architectures. As the number of amplifiers increases, the performance of both scenarios converges, with the difference eventually approaching zero. While Fig. \ref{fig:Diff_results} (a) depicts the results for a uniform power budget, Fig. \ref{fig:Diff_results} (b) represents the non-uniform case. Notably, the hybrid architecture shows a slightly larger performance delta in the uniform power budget scenario. This difference peaks when utilizing 6–8 amplifiers; however, beyond a threshold of 10 amplifiers, both architectures yield nearly identical performance. Expanding the amplifier resource allows for better power management within the SCF-based horseshoe segment despite the nonlinearity. This effectively ensures that the signal enters the HCF-based spurs with sufficient power to exploit HCF's much higher nonlinear threshold. As the number of amplifiers increases beyond 10, the power penalty of the original SCF horseshoe is neutralized by the additional amplifiers. At this point, the performance of both the hybrid and pure HCF architectures converges because the system-wide bottleneck shifts to spur segments.

\vspace{-2mm}

\begin{figure}[htb]
  \centering
\centering\includegraphics[width=0.9\linewidth]{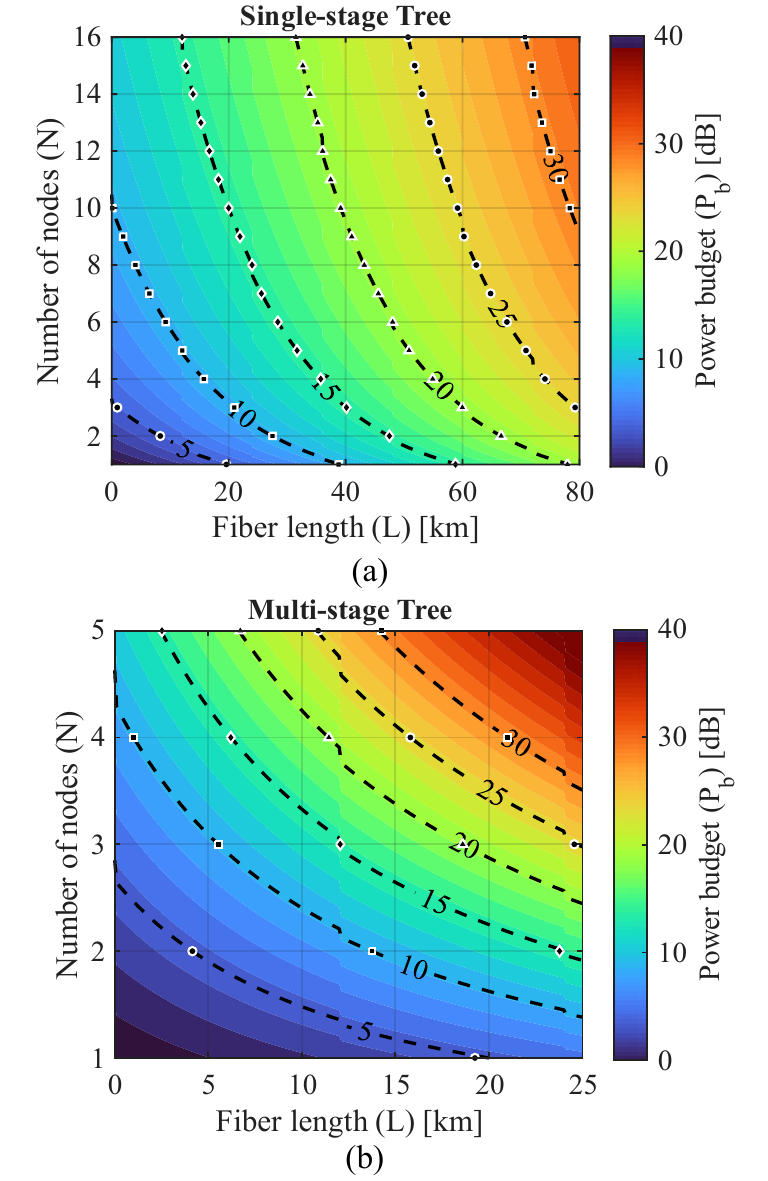}
\caption{ Number of nodes per spur, length and power budget trade-offs for (a) single-stage tree, and (b) multi-stage tree.}
\label{fig:spurpower}
\end{figure}

To quantify the trade-off between reach and power budget, we analyze the required power budget, denoted as $P_b$ [dB] using Eq.~(\ref{eq:Pb_single}) and (\ref{eq:Pb_multi}). This metric is calculated as a function of fiber length $L$ [km] and node count $N$. Figure \ref{fig:spurpower} presents a comparative representation of the required optical power budget as a function of fiber length ($L$) and number of nodes ($N$). Fig.~\ref{fig:spurpower} (a) illustrates the Single-stage tree, which demonstrates robust scaling characteristics; the widely spaced iso-power contours indicate a gradual increase in the required budget when length or number of nodes increases, maintaining manageable levels ($< 25$ dB) even as the reach extends to 80 km and the node count reaches $N=16$. In contrast, the Multi-stage Tree, depicted in Fig.~\ref{fig:spurpower} (b), exhibits a severe scalability penalty. Particularly, the power budget increases rapidly over a much shorter range ($L \le 25$ km) and for fewer nodes ($N \le 5$), evidenced by the steep gradient and dense contour lines. This highlights the cumulative impact of cascaded insertion losses, confirming that multi-stage architectures are significantly more power-limited than single-stage topologies. The adjustment of coupler ratios provides a critical degree of freedom in network design, enabling a more uniform distribution of power and improving the system's reach and scalability.

\vspace{-2mm}

To evaluate the economic feasibility of this architecture, we conduct a techno-economic break-even analysis that balances the cost premium of HCF against the capital and operational savings from reduced active amplification. As illustrated in Fig.~\ref{fig:spurpower}, the total required fiber length depends heavily on the specific architecture of the access spurs, including branching factors and the spatial distribution of endpoints. Following standard economic methodologies for emerging specialty fibers~\cite{downie2023estimating}, we assume a conservative, forward-looking HCF cost premium of $\Delta C_{\text{fiber}} = \$50/\text{km}$ per strand and a Total Cost of Ownership (TCO) per optical amplifier of $C_{\text{amp}} = \$5{,}000$.

To establish practical viability, we benchmark a break-even scenario based on power-budget parity. Our numerical results indicate that a highly optimized standard SCF baseline requires 8 to 9 amplifiers to approach its maximum achievable power budget (saturating at roughly $15.8\text{ dB}$). In contrast, the proposed HCF-enabled architecture achieves this exact performance baseline utilizing as few as 6 amplifiers---a reduction of up to 3 active units. This equipment reduction can directly offset the higher initial deployment costs of HCF. For instance, consider a targeted hybrid HCF-SCF deployment requiring $L_{\text{HCF}} = 300\text{ km}$ of HCF on the access spurs to overcome branching losses. The number of eliminated amplifiers ($N_{\text{save}}$) required to break even against the network-wide fiber premium can be calculated as follows:

\begin{equation}
    N_{\text{save}} = \frac{L_{\text{HCF}} \times \Delta C_{\text{fiber}}}{C_{\text{amp}}} = \frac{300\text{ km} \times \$50/\text{km}}{\$5{,}000} = 3
\end{equation}

By saving exactly 3 amplifiers, an operator completely absorbs the $\$15{,}000$ HCF premium simply by designing for parity with legacy SCF systems. Crucially, beyond this break-even point, any additional amplifier investments in the HCF network unlock power budgets up to $20\text{ dB}$ higher than what is physically possible in conventional SCF architectures. This demonstrates that targeted HCF edge deployments are both technically superior and economically justified, though operators must note that the required break-even threshold will scale proportionally in denser or highly distributed spur architectures that necessitate a larger HCF footprint.

Several significant challenges must be addressed before HCF can see widespread deployment in commercial settings, including cost, manufacturability, supplier diversity, and operational considerations such as testing, splicing, connectors, repairs, and integration with conventional silica fibers, and mitigating early-generation impairments such as high Polarization Mode Dispersion (PMD). Despite these hurdles, this work and others are providing early evidence that the high initial costs of deploying HCF can potentially be offset by the overall savings and performance they enable (e.g., reduced number of amplifiers, lower latency, lower power consumption).

\section{Conclusion}
This study shows that integrating HCF with DSCM transceivers transforms horseshoe-and-spur optical network design. By overcoming the nonlinearity limits of traditional SCF, this architecture doubles network reach and significantly increases the spur power budget. Because HCF is constrained mainly by amplifier output power, not fiber nonlinearity, it supports much higher per-channel power. However, increasing amplifier density eventually yields diminishing returns. Thus, carefully placing amplifiers and optimizing unbalanced coupler ratios are crucial to maximize efficiency and offset HCF’s higher upfront cost.

\small
\section*{Acknowledgments} The authors acknowledge support from the EU Horizon Europe projects ALLEGRO (GA No. 101092766) and NESTOR (GA No. 101119983), national funds through FCT – Fundação para a Ciência e a Tecnologia, I.P., and, when eligible, co-funded by EU funds under project/support UID/50008/2025 – Instituto de Telecomunicações, with DOI identifier https://doi.org/10.54499/UID/50008/2025.
\normalsize
% \red{we might need to add more article on PROTEUS-6G - i will change it later.}

% Patch the bibliography to use footnotesize

% \bibliography{sample.bib}

\bibliographystyle{ieeetr}
\bibliography{sample}

\end{document}